\documentclass[prd,aps]{revtex4}
\usepackage[T2A]{fontenc}
\usepackage{amsmath,amssymb}
\usepackage{eucal}
\usepackage{color}
\usepackage[dviwindo]{graphicx}
\newcommand{\bs}{\begin{subequations}}
\newcommand{\es}{\end{subequations}}
\numberwithin{equation}{section}
\newcommand{\ben}{\begin{eqnarray}}
\newcommand{\een}{\end{eqnarray}}
\newcommand{\la}{\label}

\begin{document}

%\DeclareGraphicsExtensions{.jpg,.pdf,.mps,.png}

\title{A new approach to the connection problem for local solutions to the general Heun equation}

\author{P. P.  Fiziev
\footnote{fizev@phys.uni-sofia.bg\,\,\,and\,\,\,
fizev@theor.jinr.ru}}
\affiliation{Sofia University Foundation for Theoretical and Computational Physics and Astrophysics, Boulevard
5 James Bourchier, Sofia 1164, Bulgaria\\
and\\
BLTF, JINR, Dubna, 141980 Moscow Region, Rusia}

 \begin{abstract}
We present new solution of the the connection problem for local solutions to the general Heun equation.
Our approach is based on the symmetric form of the Heun's differential equation \cite{Fiziev14,Fiziev16}
with four different regular singular points $z_{1,2,3,4}$.
The four special regular points in the complex plane: $Z_{123},Z_{234},Z_{341},Z_{412}$ are the centers of the
circles, defined by the different triplets $\{z_k,z_l,z_m\}$ with the corresponding different indices and play a fundamental role
since the coefficients of the connection matrix can be expressed using the values of local solutions
of the general Heun's equation at these points.
A special case when all coefficients can be calculated using only one of the points $Z_{klm}$ is also considered.

\vskip .2truecm
MSC classification scheme numbers: 34A25, 34B30, 11B37
\end{abstract}
%%%%%%%%%%%%%%%%%%%%%%%%%%%%%%%%%%%%%%%%%%%%%%%%%%%%%%%%%%
%\draft
\sloppy
%\scrollmode
%%%%%%%%%%%%%%%%%%%%%%%%%%%%%%%%%%%%%%%%%%%%%%%%%%%%%%%%%%%
\maketitle

\section{Introduction}

\begin{figure}[ht!]
\vskip -0.truecm
\hskip -7.truecm
\begin{minipage}{6.5cm}
\includegraphics[width=.7\textwidth]{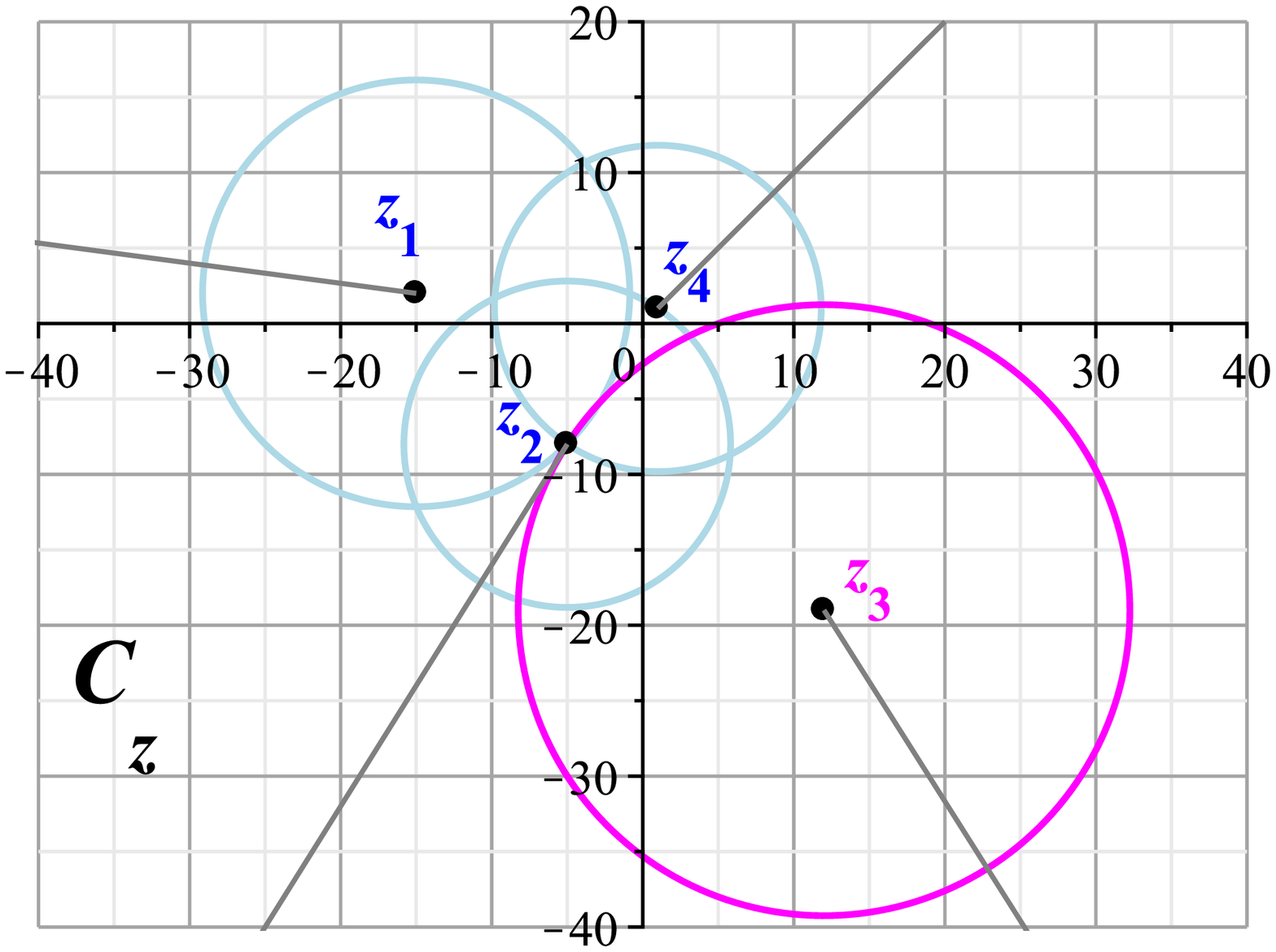}
\vskip .3truecm
\caption{The singular points $z_{1,2,3,4}\in\mathbb{\tilde C}_z$ of Eq.\eqref{dF},
the corresponding circles of convergence and branching cuts going to $\infty$.}
\label{fig1}
\end{minipage}
\vskip -5.9truecm
\hskip 7.truecm
\begin{minipage}{6.5cm}
\includegraphics[width=.7\textwidth]{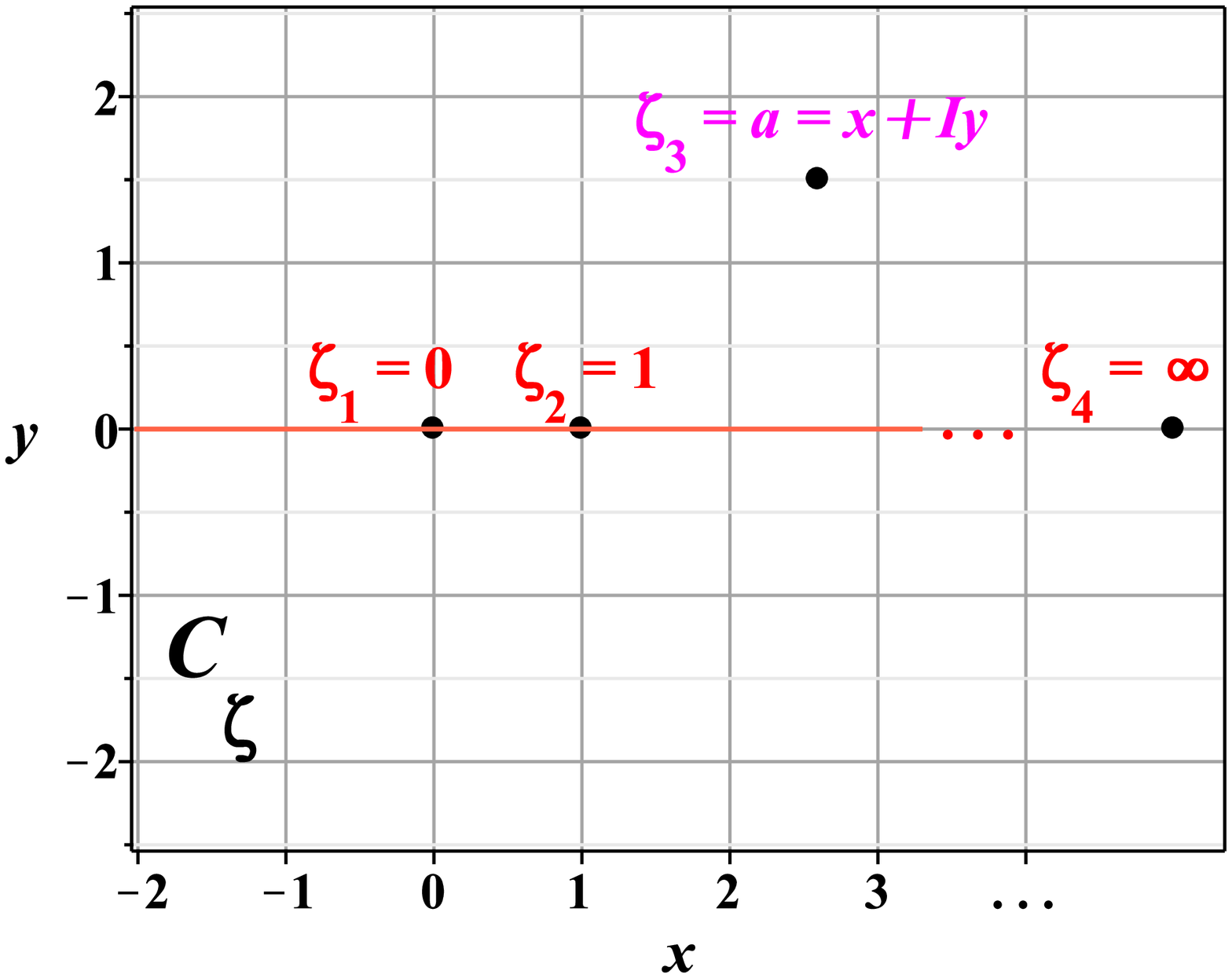}
\vskip .6truecm
\caption{The singular points $\zeta_{1,2,3,4}\in\mathbb{\tilde C}_\zeta$ of  Eq.\eqref{dHeunG}.
The dependence on indices prevents a common treatment of branching points.}
\label{fig2}
\end{minipage}
\end{figure}

A new form of the general Heun's equation for the function $\mathcal{F}(z)$ defined on the complex plane $\mathbb{\tilde C}$:
\ben
\mathcal{F}^{\prime\prime}+{\frac 1 2}\left(\sum_{j=1}^4{\frac 1{z-z_j}}\right)\mathcal{F}^\prime+
{\frac 1 {P(z)}\left(\lambda+\sum_{j=1}^4{\frac{q_j}{z-z_j}}\right)}\mathcal{F} = 0
\la{dF}
\een
was introduced and studied in \cite{Fiziev14,Fiziev16}.
Everywhere the prime denotes differentiation with respect to an independent variable.
Equations \eqref{dF} and the function $\mathcal{F}(z)$ have four regular singular points
$z_{1,2,3,4} \neq 0$, eigenvalue $\lambda$,
$P(z)=\prod_{j=1}^4(z-z_j)$, $q_j=\alpha_j\beta_jP^\prime(z_j)$, and
indices $\alpha_j={\frac 1 2}\cos(\chi_j)^2$, $\beta_j={\frac 1 2}\sin(\chi_j)^2$ where $\chi_{1,2,3,4}\in\mathbb{\tilde C}$ are free parameters.
Since  $\alpha_j+ \beta_j ={\frac 1 2}$, the singular points $z_{1,2,3,4}$ are in any case branching points, see Fig.\ref{fig1}:
the grey lines present the branching cuts.
\hskip -3.truecm

Usually, the general Heun's equation is written in the most popular at present form  \cite{Heun,Ron,SL,NIST}:
\ben
\hskip -.truecm
H''+\left({\frac \gamma \zeta}+{\frac{\delta}{\zeta-1}}+{\frac{\epsilon}{\zeta-a}}\right) H'+
{\frac{\alpha\beta \zeta-\lambda}{\zeta(\zeta-1)(\zeta-a)}}H = 0,\,\,\,\,\,
\gamma\!+\!\delta\!+\!\epsilon\!=\!\alpha\!+\!\beta\!+\!1.
\hskip 1truecm
\la{dHeunG}
\een
In contrast to Eq. \eqref{dF},  Eq. \eqref{dHeunG} is not symmetric with respect to the four regular singular points
$\zeta_{1,2,3,4}= 0, 1, a, \infty$ (See Fig.\ref{fig2}) with indices
$\{ 0, 1-\gamma\},\, \{ 0, 1-\delta \},\, \{ 0, \gamma+\delta-\alpha-\beta \},\, \{ \alpha, \beta \}$.

Note that at present the Riemannian surfaces of the Heun's functions are usually ignored
and the branch cuts, as a rule, are not justified.
Such a careless approach leads to difficulties, especially in the numerical
calculations, since it leads to branch mishmash. A good exception are the recent papers \cite{Motygin15a,Motygin15b}.

The local solutions of Eqs. \eqref{dF}, Eq. \eqref{dHeunG} can be easily obtained in the form of the Frobenius series expansions \cite{CL}
with the coefficients defined by the corresponding recurrence relations: three-term ones for Eq. \eqref{dHeunG}, see \cite{Ron,SL,NIST},
and nine-term onesa - for Eq. \eqref{dF}, see \cite{Fiziev14,Fiziev16}.

A long standing connection problem is how to express any of the pears of linear-independent local solutions of  Eq. \eqref{dHeunG}
around some singular point $\zeta_{1,2,3,4}$ as a linear combination of any one of the other three such pairs.
The existing attempts to solve this basic problem and the corresponding not very satisfactory results are described in \cite{Ron,SL,NIST},
and in \cite{Schafke80a,Schafke80b,Schafke84}.

Here we give a simple resolution of this problem using the solutions of Eq. \eqref{dF} and their relation with the solutions of Eq. \eqref{dHeunG}.

\section{Some auxiliary results}

\subsection{The center and the radius of a circle through three different points $z_{k,l,m} \in \mathbb{\tilde C}$}\label{Cklm}
{\em {\bf \em Lemma 1}:  Let $z_{k,l,m} \in \mathbb{\tilde C}$ be any three different points $z_k\neq z_l \neq z_m \neq z_k$
($k,l,m=1,2,3,4$).
Then the center $Z_{klm}$ and the radius $R_{klm}$ of the circle through these three points are given by the following expressions:}
\ben
Z_{klm}={\frac{\bar z_k z_k(z_l-z_m)+\bar z_l z_l(z_m-z_k)+\bar z_m z_m(z_k-z_l)}
{\bar z_k(z_l-z_m)+\bar z_l(z_m-z_k)+\bar z_m(z_k-z_l)}},
\la{center}
\een
and
\ben
R_{klm}= {\frac {|z_k-z_l||z_l-z_3||z_3-z_k|}
{|\bar z_k(z_l-z_m)+\bar z_l(z_m-z_k)+\bar z_m(z_k-z_l)|}}.
\la{radius}
\een

Here the bar denotes complex conjugation and |z| is the modulus of the corresponding complex number $z$.
Note that $Z_{klm}$ and $R_{klm}$ are not analytic functions of the variables $z_{k,l,m}$, since they depend also on $\bar z_{k,l,m}$.

The proof is a simple application of elementary geometry and the definitions of the center $Z_{klm}$
as a crossing point of the two complex straight lines $Z_k(\lambda_1)={\frac {z_k+z_l} 2} +i(z_l-z_k)\lambda_1$ and
$Z_l(\lambda_2)={\frac {z_m+z_l} 2} +i(z_m-z_l)\lambda_2$ with real parameters $\lambda_{1,2} \in \mathbb{R}$.

\begin{figure}[ht!]
\vskip -0.truecm
\hskip -0.truecm
\begin{minipage}{6.5cm}
\includegraphics[width=.7\textwidth]{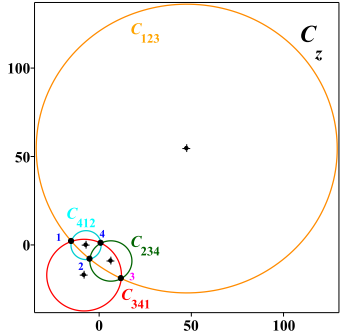}
\vskip .5truecm
\caption{The circles $C_{123},C_{234},C_{341},C_{412}$.}
\label{fig3}
\end{minipage}
\end{figure}

Since between the singular points of Eq. \eqref{dF} $z_{1,2,3,4}$ one can choose four triples of points $z_{1,2,3},z_{2,3,4},z_{3,4,1},z_{4,1,2}$,
according to the above Lemma 1 we obtain four circles $C_{klm}$, see Fig.\ref{fig3}.
%,
\subsection{The different Moebius mappings needed for our purposes}\label{Moebius}

{\em {\bf \em Lemma 2}: Any three different points $z_{k,l,m}\in \mathbb{\tilde C}$ can be mapped onto any three different points $\zeta_{k,l,m} \in \mathbb{\tilde C}$ by
proper Moebius transformation:}
\ben
z\leftrightarrow \zeta: \quad \zeta={\frac {Az+B}{Cz+D}}\leftrightarrow z={\frac {D\zeta-B}{-C\zeta+A}},\quad \forall z,\zeta \in \mathbb{\tilde C}, \quad AD-BC\neq 0.
\la{Mobius}
\een
For example,  the explicit formulas for the case $z_{k,l,m}=z_{1,2,4} $ are
\ben
A=\begin{vmatrix}
z_1\zeta_1,&\zeta_1,&1 \\
z_2\zeta_2,&\zeta_2,&1 \\
z_4\zeta_4,&\zeta_4,&1 \\
\end{vmatrix}, \quad
B=\begin{vmatrix}
z_1\zeta_1,&z_1,&\zeta_1 \\
z_2\zeta_2,&z_2,&\zeta_2 \\
z_4\zeta_4,&z_4,&\zeta_4 \\
\end{vmatrix}, \quad
C=\begin{vmatrix}
z_1,&\zeta_1,&1 \\
z_2,&\zeta_2,&1 \\
z_4,&\zeta_4,&1 \\
\end{vmatrix}, \quad
D=\begin{vmatrix}
z_1\zeta_1,&z_1,&1 \\
z_2\zeta_2,&z_2,&1 \\
z_4\zeta_4,&z_4,&1 \\
\end{vmatrix}.
\la{z_zeta_det}
\een

{\em {\bf \em Lemma 3}:  The Mobius transformation
\ben
z \mapsto \zeta(z;z_1,z_2,z_4) = {\frac {z_2-z_4}{z_2-z_1}}{\frac {z-z_1}{z-z_4}}
\la{z_zeta}
\een
with parameters $z_1,z_2,z_4$ maps the four arbitrary different points
$\{z_1,z_2,z_3,z_4\} \in \mathbb{\tilde C}$ onto the points $\{\zeta_1,\zeta_2,\zeta_3,\zeta_4,\}=\{0,1,a,\infty \}\in \mathbb{\tilde C}$,
thus describing the relation between the singular points of solutions to Eqs. \eqref{dF} and \eqref{dHeunG}.}

Here
\ben
a=a(z_1,z_2,z_3,z_4)= {\frac {z_2-z_4}{z_2-z_1}}{\frac {z_3-z_1}{z_3-z_4}}= {\frac {(\zeta_2-\zeta_4)(\zeta_3-\zeta_1)}{(\zeta_2-\zeta_1)(\zeta_3-\zeta_4)}}
= \text{invariant} =\zeta(z_3)= \zeta_3
\la{a}
\een
is the invariant of the Mobius transformation, called also the cross-ratio (see, for example, \cite{Golubev,AF}.

The origin $z=0$ will play in our consideration a very special role.
The Mobius transformation \eqref{z_zeta} maps it onto the regular point
\ben
z=0 \mapsto \zeta_0 =\zeta(0;z_1,z_2,z_4)= {\frac {z_2-z_4}{z_2-z_1}}{\frac {z_1}{z_4}}.
\la{0_zeta}
\een

The inverse Mobius transformation is
\ben
\zeta \mapsto z(\zeta;z_1,z_2,z_4)  = {\frac {z_4(z_2-z_1)\zeta - z_1(z_2-z_4)}{(z_2-z_1)\zeta - (z_2-z_4)}}.
\la{zeta_z}
\een

{\em {\bf \em Consequence 1}:} The solutions of Eqs. \eqref{dF}  and \eqref{dHeunG} are simply related. For example,
using the Maple notation for the general Heun's functions and Eqs. \eqref{z_zeta} and \eqref{a} we can write down
\ben
\mathcal{F}(z;z_1,z_2,z_3,z_4;q_1,q_2,q_3,q_4;\lambda)&=&HeunG(a;q;\alpha,\beta,\gamma,\delta, \zeta)\prod_{k=1}^4 (z-z_k)^{\nu_k},\\
\nu_1=-\alpha_2-\alpha_3-\alpha_4,\nu_2&=&\alpha_2,\quad \nu_3=\alpha_3,\quad\nu_4=\alpha_4.\nonumber
\la{FH}
\een

{\em {\bf \em Consequence 2}:} Using formulae \eqref{z_zeta_det} we see that any three different points $z_{k,l,m}\in \mathbb{\tilde C}$ can be mapped
on three different points on the unit circle: $z_k=e^{i\Phi_k},\,\,z_l=e^{i\Phi_l},\,\,z_m=e^{i\Phi_m}$, $\Phi_{k,l,m} \in \mathbb{R}$. For example,
in the case $k,l,m=1,2,4$ the corresponding coefficients are
\ben A&=&(z_1-z_2)e^{i(\Phi_1+\Phi_2)}-(z_1-z_4)e^{i(\Phi_1+\Phi_4)}+(z_2-z_4)e^{i(\Phi_2+\Phi_4)},\nonumber\\
B&=&-z_4(z_1-z_2)e^{i(\Phi_1+\Phi_2)}+z_2(z_1-z_4)e^{i(\Phi_1+\Phi_4)}-z_1(z_2-z_4)e^{i(\Phi_2+\Phi_4)},\nonumber\\
C&=&-(z_1-z_2)e^{i\Phi_4}+(z_1-z_4)e^{i\Phi_2}-(z_2-z_4)e^{i\Phi_1},\nonumber\\
D&=&z_4(z_1-z_2)e^{i\Phi_4}-z_2(z_1-z_4)e^{i\Phi_2}+z_1(z_2-z_4)e^{i\Phi_1}.
\la{ABCD}
\een

{\em {\bf \em Consequence 3}:} From the above formulas we obtain the Moebius transformation
\ben
\zeta_{1,2,3,4}= 0, 1, \infty \rightleftarrows z_{1,2,4}=e^{i\Phi_1},e^{i\Phi_2},e^{i\Phi_4}
\la{M124}
\een
in the form
\ben
\zeta \mapsto z(\zeta;\Phi_1,\Phi_2,\Phi_4)=e^{i\Phi_4}{\frac {\zeta-\zeta_0 }{\zeta-\overline{\zeta_0}}},\quad
z \mapsto \zeta(z;\Phi_1,\Phi_2,\Phi_4)=\overline{\zeta_0}\,{\frac{z-e^{i\Phi_1}}{z-e^{i\Phi_4}}},
\la{MPhi124}
\een
where we introduce $\zeta_0=x_0+iy_0$ (see Eq. \eqref{0_zeta}.) with $x_0=c_{14}s_{42}/s_{12}$, $y_0=s_{14}s_{42}/s_{12}$, $s_{kl}=
\sin({\frac{\Phi_k-\Phi_l}{2}})$, and $c_{kl}=\cos({\frac{\Phi_k-\Phi_l}{2}})$.

\section{Solution of the connection problem}\label{SolCP}

\subsection{The proper choice of the real angles $\Phi_{1,2,4} \in \mathbb{R}$ in $z_1=e^{i\Phi_1},\,\,z_2=e^{i\Phi_2},\,\,z_4=e^{i\Phi_4}$ }\label{Phi124}
\begin{figure}[ht!]
\vskip -0.truecm
\hskip -5.truecm
\begin{minipage}{6.cm}
\includegraphics[width=.7\textwidth]{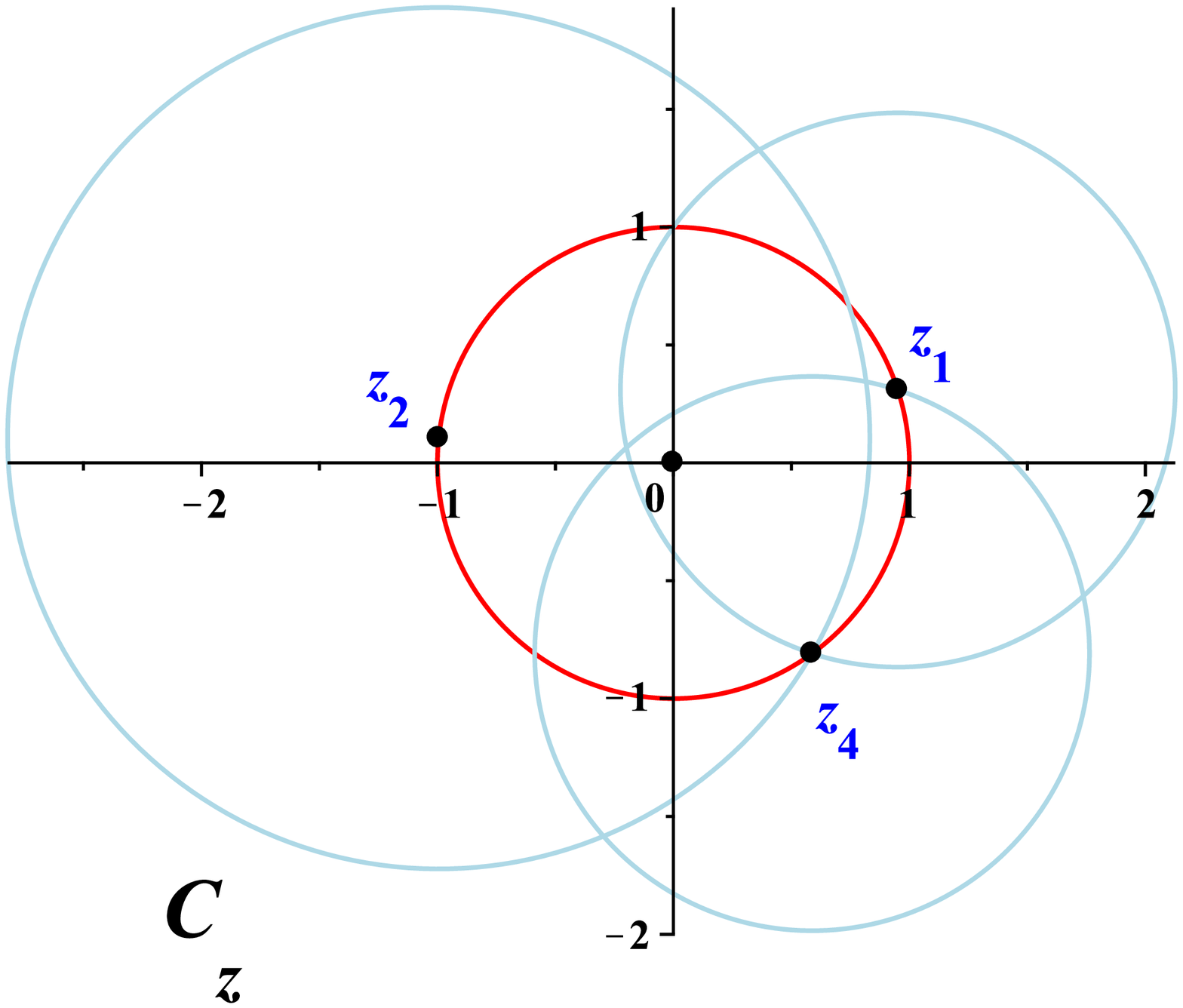}
\vskip .5truecm
\caption{A choice of $\Phi_{1,2,4}$ for which \\
$z=0 \in D_1\cap D_2\cap D_4$.}
\label{fig4}
\end{minipage}
\vskip -5.4truecm
\hskip 8.truecm
\begin{minipage}{6.cm}
\includegraphics[width=.6\textwidth]{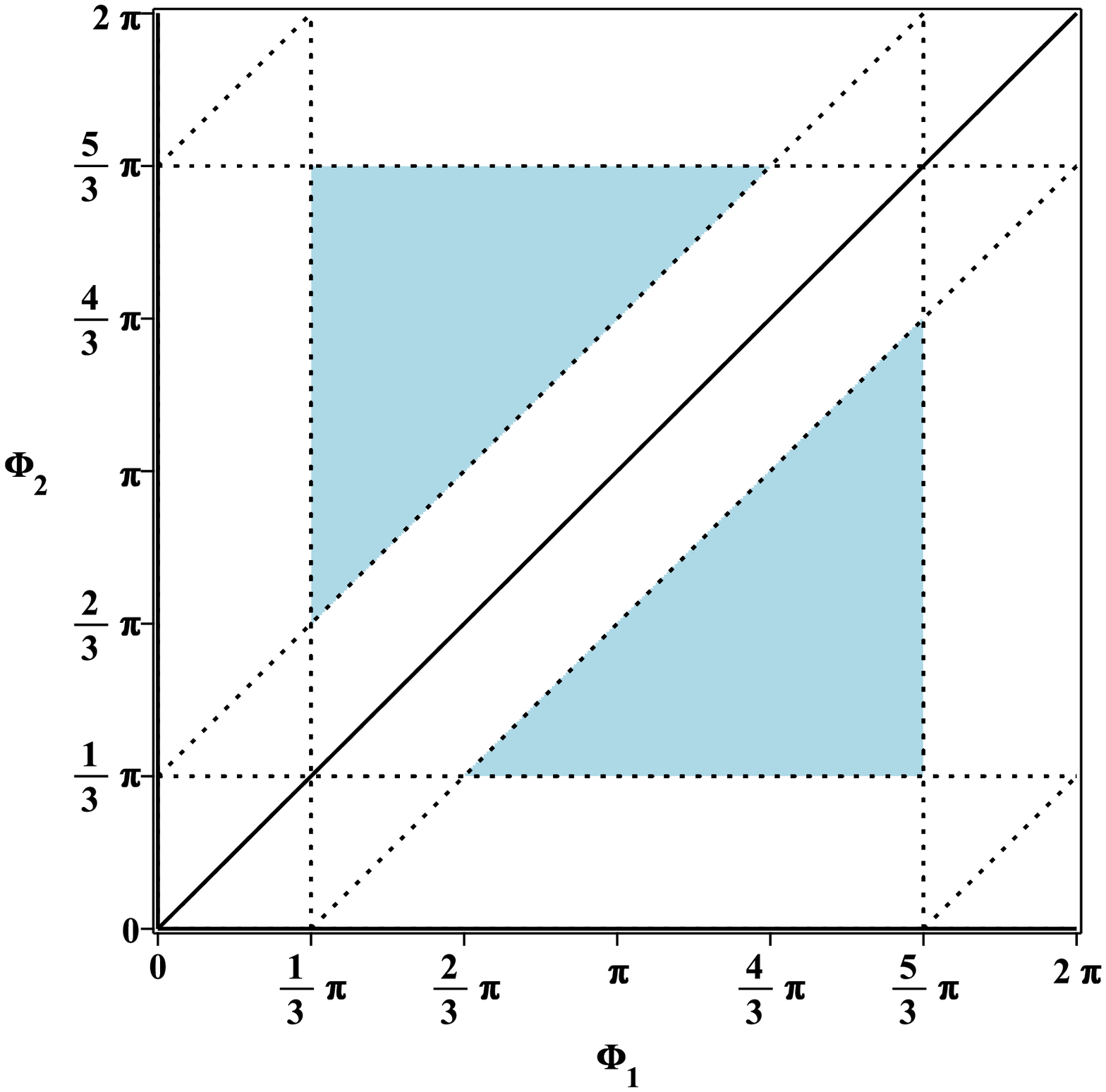}
\vskip .6truecm
\caption{The domain  of $\Phi_{1,2}$ (mod $2\pi$) \\for which $z=0 \in D_1\cap D_2\cap D_4$.}
\label{fig5}
\end{minipage}
\end{figure}

Using the Moebius mapping \eqref{MPhi124} we can solve the connection problem under a proper choice of the real angles $\Phi_{1,2,4}$.
The two possible situations are shown:
\begin{enumerate}
\item  In Fig.\ref{fig1} where $z=0 \notin D_1\cap D_2\cap D_4$, and
\item  In Fig.\ref{fig4} where $z=0 \in D_0\cap D_1\cap D_2\cap D_4$.
\end{enumerate}
Here $D_{1,2,4}$ denote the discs of convergence of the Frobenius series of the local solutions around $z_{1,2,4}$,
$D_0$ is the unite disc of convergence of the Taylor series of the local solutions around $z=0$.

To have  $z=0 \in D_1\cap D_2\cap D_4$ for $z_1=e^{i\Phi_1},\,\,z_2=e^{i\Phi_2},\,\,z_4=e^{i\Phi_4}$, $\Phi_{1,2,4} \in \mathbb{R}$
we must superimpose on the three dimensional torus $\mathbb{T}^{\,\,\,\,(3)}_{\Phi_1\Phi_2\Phi_4}$ the following:
\vskip .2truecm

{\em {\bf \em Condition A}:}
\ben
r_{12}=2\left|\sin\left({\tfrac {\Phi_1-\Phi_2}{2}}\right)\right|>1,\, r_{24}=2\left|\sin\left({\tfrac {\Phi_2-\Phi_4}{2}}\right)\right|>1,\,
r_{41}=2\left|\sin\left({\tfrac {\Phi_4-\Phi_1}{2}}\right)\right|>1.
\la{CondA}
\een
Here $r_{kl}=|z_k-z_l|$.
Because of the translational invariance: $\forall \phi$ the substitution
$\Phi_1 \to \Phi_1+\phi, \Phi_2 \to \Phi_2+\phi, \Phi_3 \to \Phi_3+\phi$ does not change the Condition A,
we can fix the value of the angle $\Phi_4$, say $\Phi_4=0$.
Then the graphical solution of the Condition A reduces to a solution
on the torus $\mathbb{T}^{\,\,\,\,(2)}_{\Phi_1\Phi_2}$
and is shown in Fig.\ref{fig5}.

Obviously, we are free to choose any point $\{\Phi_1,\Phi_2\}$ in the two-component light-blue-domain shown in Fig.\ref{fig5} and thus to ensure
the fulfilment of the condition $z=0 \in D_1\cap D_2\cap D_4$.

\subsection{Connection matrices}

Now we are ready to prove the following:
\vskip .2truecm
{\em {\bf \em Proposition 1}:
\vskip .2truecm
Under Condition A the connection matrix
\ben
 \hat{C}(z_k,z_l)
=\begin{Vmatrix}
{C}_{11}(z_k;z_l) & {C}_{12}(z_k;z_l)\\
{C}_{21}(z_k;z_l) & {C}_{22}(z_k;z_l)
\end{Vmatrix}
\la{Ckl}
\een
for the local solutions $\mathcal{F}_{1,2}(z;z_{k,l,m})$ of Eq. \eqref{dF} around any pair of singular points $z_k,z_l$
can be expressed using their values at the common regular point $z=0 \in D_1\cap D_2\cap D_4$ and has the following matrix elements:
\begin{subequations}\label{C:1,2,3,4}
\ben
 {C}_{11}(z_k;z_l)={\frac {\mathcal{F}_1(0;z_k)\mathcal{F}_2(0;z_l)^\prime -\mathcal{F}_1(0;z_k)^\prime\mathcal{F}_2(0;z_l) }
                    {\mathcal{F}_1(0;z_l)\mathcal{F}_2(0;z_l)^\prime -\mathcal{F}_1(0;z_l)^\prime\mathcal{F}_2(0;z_l) }},\label{C:1}\\
{C}_{12}(z_k;z_l)=-{\frac {\mathcal{F}_1(0;z_k)\mathcal{F}_1(0;z_l)^\prime -\mathcal{F}_1(0;z_k)^\prime\mathcal{F}_1(0;z_l) }
                    {\mathcal{F}_1(0;z_l)\mathcal{F}_2(0;z_l)^\prime -\mathcal{F}_1(0;z_l)^\prime\mathcal{F}_2(0;z_l) }} \label{C:2}\\
{C}_{21}(z_k;z_l)={\frac {\mathcal{F}_2(0;z_k)\mathcal{F}_2(0;z_l)^\prime -\mathcal{F}_2(0;z_k)^\prime\mathcal{F}_2(0;z_l) }
                    {\mathcal{F}_1(0;z_l)\mathcal{F}_2(0;z_l)^\prime -\mathcal{F}_1(0;z_l)^\prime\mathcal{F}_2(0;z_l) }},\label{C:3}\\
{C}_{22}(z_k;z_l)=-{\frac {\mathcal{F}_2(0;z_k)\mathcal{F}_1(0;z_l)^\prime -\mathcal{F}_2(0;z_k)^\prime\mathcal{F}_1(0;z_l) }
                    {\mathcal{F}_1(0;z_l)\mathcal{F}_2(0;z_l)^\prime -\mathcal{F}_1(0;z_l)^\prime\mathcal{F}_2(0;z_l) }}.\label{C:4}
 \la{Ckl_elements}
\een
\end{subequations}
}

{\em {\bf \em Proof}}:
\vskip .2truecm
I. Suppose the Condition A is fulfilled for the choice $z_{k,l,m}= z_{1,2,4}$ and denote by:
\vskip .2truecm
1) $ \mathcal{F}_{1,2}(z;0)$ - the pair of local solutions of Eq. \eqref{dF}
which have the Taylor series expansion around $z=0$ and obey the
conditions
\begin{subequations}\label{F0:1,2}
\ben
\mathcal{F}_1(0;0)=1,\quad \mathcal{F}_1^{\,\prime}(0;0)=0,  \la{F0:1}\\
\quad \mathcal{F}_2(0;0)=0,\quad \mathcal{F}_2^{\,\prime}(0;0)=1.\la{F0:2}
\een
\end{subequations}
These solutions are linearly independent and form the vector
\ben
\mathbf{F}(z,0)
=\begin{pmatrix}
\mathcal{F}_1(z;0) \\
\mathcal{F}_2(z;0)
\end{pmatrix}.
\la{bfF0}
\een

1) $ \mathcal{F}_{1,2}(z;z_{1,2,4})$ - the pair of local solutions of
Eq. \eqref{dF} which have the Frobenius series expansion around $z=z_{1,2,4}$ and obey the
condition  $ W[ \mathcal{F}_{1}(z;z_{1,2,4}), \mathcal{F}_{2}(z;z_{1,2,4})\neq 0$.
These solutions are linearly independent and form the vector
\ben
\mathbf{F}(z,z_{1,2,4})
=\begin{pmatrix}
\mathcal{F}_1(z;z_{1,2,4}) \\
\mathcal{F}_2(z;z_{1,2,4})
\end{pmatrix}.
\la{bfF124}
\een

Hence, there exist nonsingular constant connection matrices $\hat{C}(z_k,0)$ and $\hat{C}(z_k,z_l)$, such that
for any $z\in D_0 \cap D_1 \cap D_2 \cap D_4$
\ben
\mathbf{F}(z,z_k)&=& \hat{C}(z_k,0) \mathbf{F}(z,0)
,\,\,\mathbf{F}(z,z_k)= \hat{C}(z_k,z_l) \mathbf{F}(z,z_l) \Rightarrow \nonumber \\
 \hat{C}(z_k,z_l)&=& \hat{C}(z_k,0)\hat{C}(z_l,0)^{-1}=\hat{C}(z_k,0)\hat{C}(0,z_l).
\la{cF0Fk}
\een
The connection matrices obey the obvious chain identities
\begin{subequations}\label{Cident:1,2,3,4}
\ben
\hat{C}(z_k,z_k)&\equiv& \hat{1},\label{Cident:1}\\
\hat{C}(z_k,z_l)^{-1}&=& \hat{C}(z_l,z_k),\label{Cident:2}\\
\hat{C}(z_k,z_l)\hat{C}(z_l,z_k)&=&\hat{C}(z_k,z_l),\label{Cident:3}\\
\hat{C}(z_k,z_l)\hat{C}(z_l,z_m)&=&\hat{C}(z_k,z_m),\dots\label{Cident:4}
\een
\end{subequations}

The direct use of the above definitions and Eq. \eqref{F0:1,2} leads after some algebra to our basic result, Eqs. \eqref{Ckl}, \eqref{C:1,2,3,4}.

II. The choice $z_{k,l,m}= z_{1,2,4}$  is not unique and can be replaced by any other one:
$z_{k,l,m}= z_{2,3,4}$, $z_{k,l,m}= z_{3,4,1}$, or $z_{k,l,m}= z_{4,1,2}$ (see Fig. \ref{fig3}).
Under the last choices the relations and the formulas for transition from solutions of Eq. \eqref{dF} to the solutions of Eq. \eqref{dHeunG},
as described in Section \ref{Moebius}, may be more complicated,
but the above procedure (point I in the present Section) can be repeated with the same result
and yields formulas,  Eqs. \eqref{Ckl}, \eqref{C:1,2,3,4} with the corresponding set of indices.
Since in each pair of the sets of indies we have just two common indices (note that in Fig. \ref{fig3}
the section of any two of the circles $C_{klm}$ consists of just two singular points by construction.),
we are able to connect explicitly all local solutions $\mathbf{F}(z,z_k)$ for $k=1,2,3,4$
using the corresponding well defined values of these local solutions at all different
regular points $Z_{klm}$.

\subsection{Solution of the connection problem using only one regular point $Z_{klm}$}

An interesting point is to clarify the conditions under which it is possible
to solve the connection problem for all local solutions $\mathbf{F}(z,z_k)$ for $k=1,2,3,4$
using their well defined values at only one of the regular points $Z_{klm}$. For this purpose we have to
ensure that this point belongs simultaneously to the domains of convergence of the Frobenios series of all
solutions $\mathbf{F}(z,z_k)$.

Consider once more the simplest choice  $z_{k,l,m}= z_{1,2,4}$. For the validity of the last new acquirement
we obviously have to add to the Condition A an additional one:

{\em {\bf \em Condition B}:}
\ben
\min\{r_{13},r_{23},r_{43}\} > r_3,\quad \text{where}\quad r_3=|z_3|.
\la{CondB}
\een
The condition B is much more complicated than the Condition A and depends on the values of the parameter a.
In the present case, $a$ acquires the form, which follows from Eq. \eqref{MPhi124}:
\ben
a(\zeta;\Phi_1,\Phi_2,\Phi_4)=e^{i\Phi_4}{\frac {\zeta_3-\zeta_0 }{\zeta_3-\overline{\zeta_0}}}\,.
\la{a124}
 \een

\begin{figure}[ht!]
\vskip -0.truecm
\hskip -5.truecm
\begin{minipage}{6.cm}
\includegraphics[width=.7\textwidth]{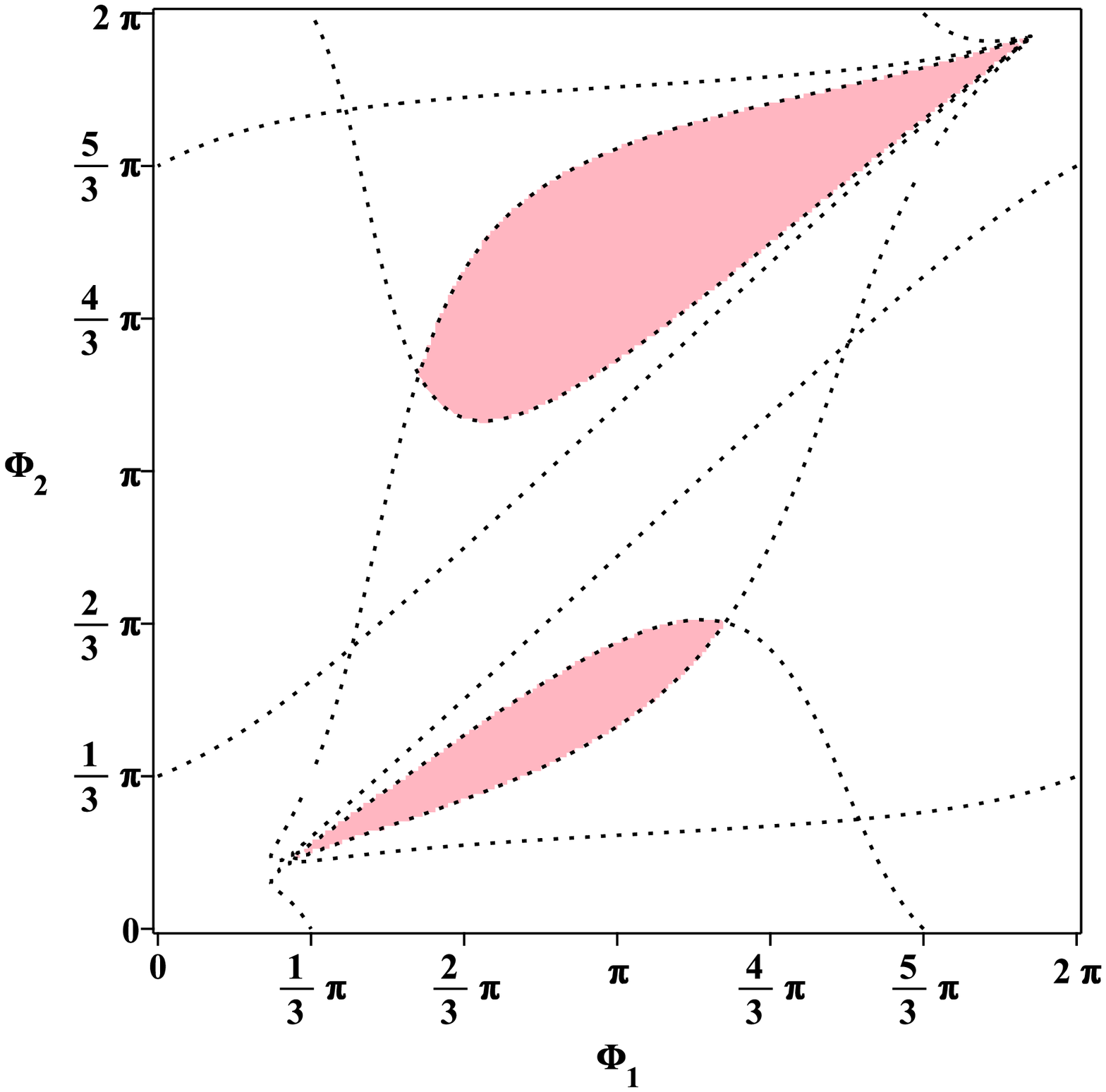}
\vskip .5truecm
\caption{The domain $\Phi_{1,2,4}$ with valid Condition B .}
\label{fig6}
\end{minipage}
\vskip -5.9truecm
\hskip 8.truecm
\begin{minipage}{6.cm}
\includegraphics[width=.7\textwidth]{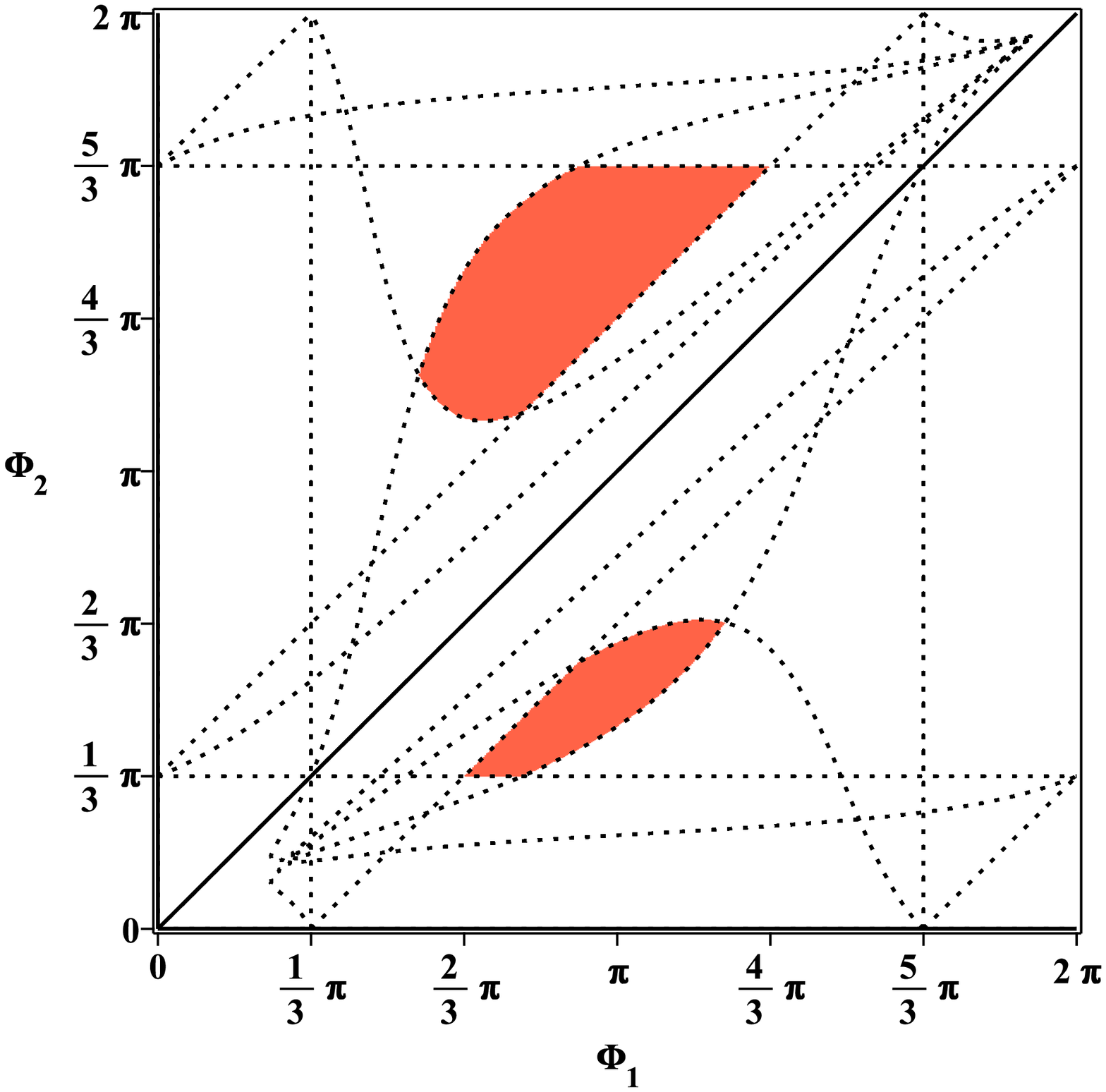}
\vskip .5truecm
\caption{The domain  of $\Phi_{1,2}$ (mod $2\pi$) with simultaneously valid Conditions A and B.}
\label{fig7}
\end{minipage}
\end{figure}

\begin{figure}[ht!]
\vskip -0.truecm
\hskip -5.truecm
\begin{minipage}{6.cm}
\includegraphics[width=.7\textwidth]{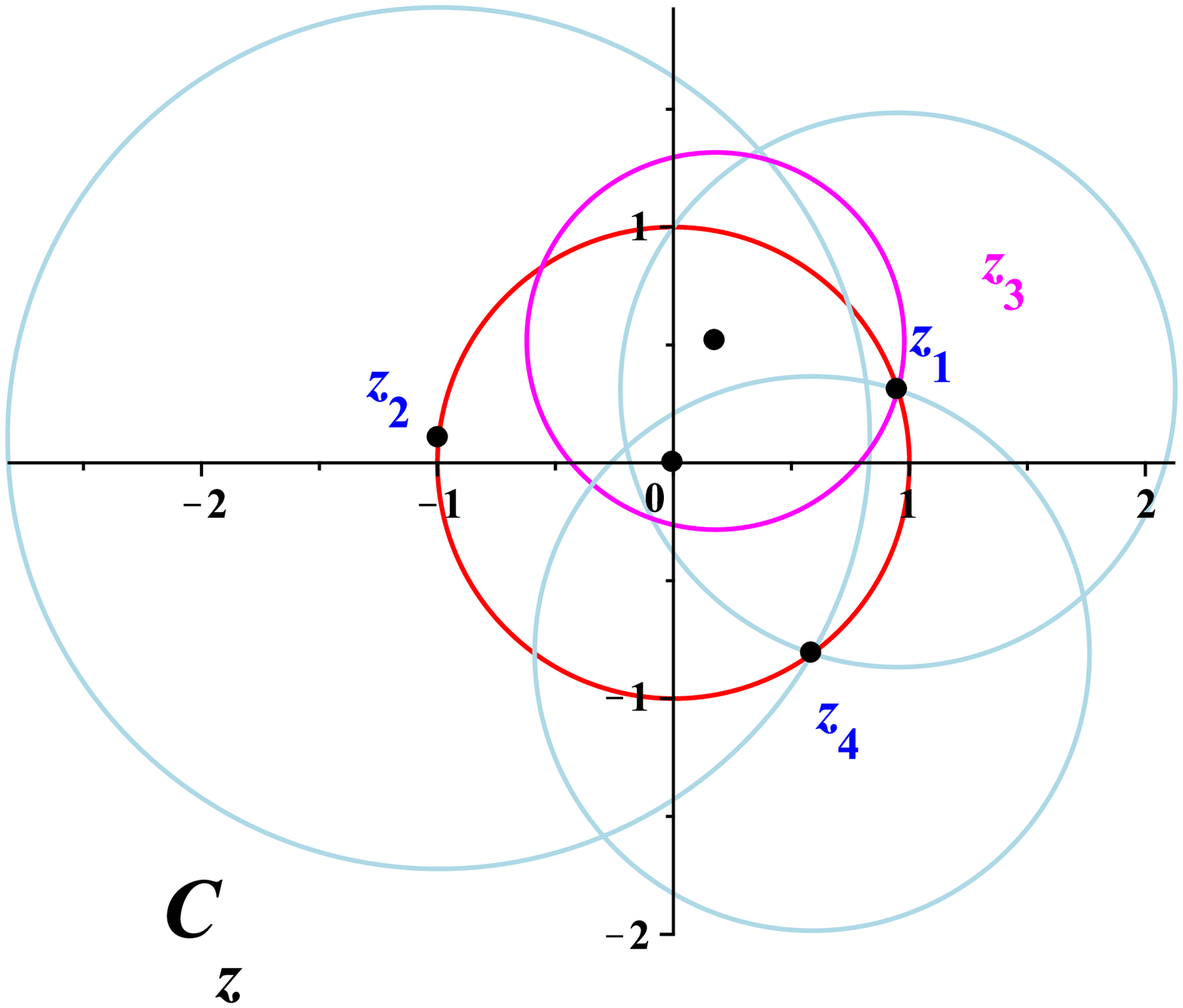}
\vskip .5truecm
\caption{An example of a configuration with simultaneously valid Conditions A and B.}
\label{fig8}
\end{minipage}
\vskip -5.7truecm
\hskip 8.truecm
\begin{minipage}{6.cm}
\includegraphics[width=.7\textwidth]{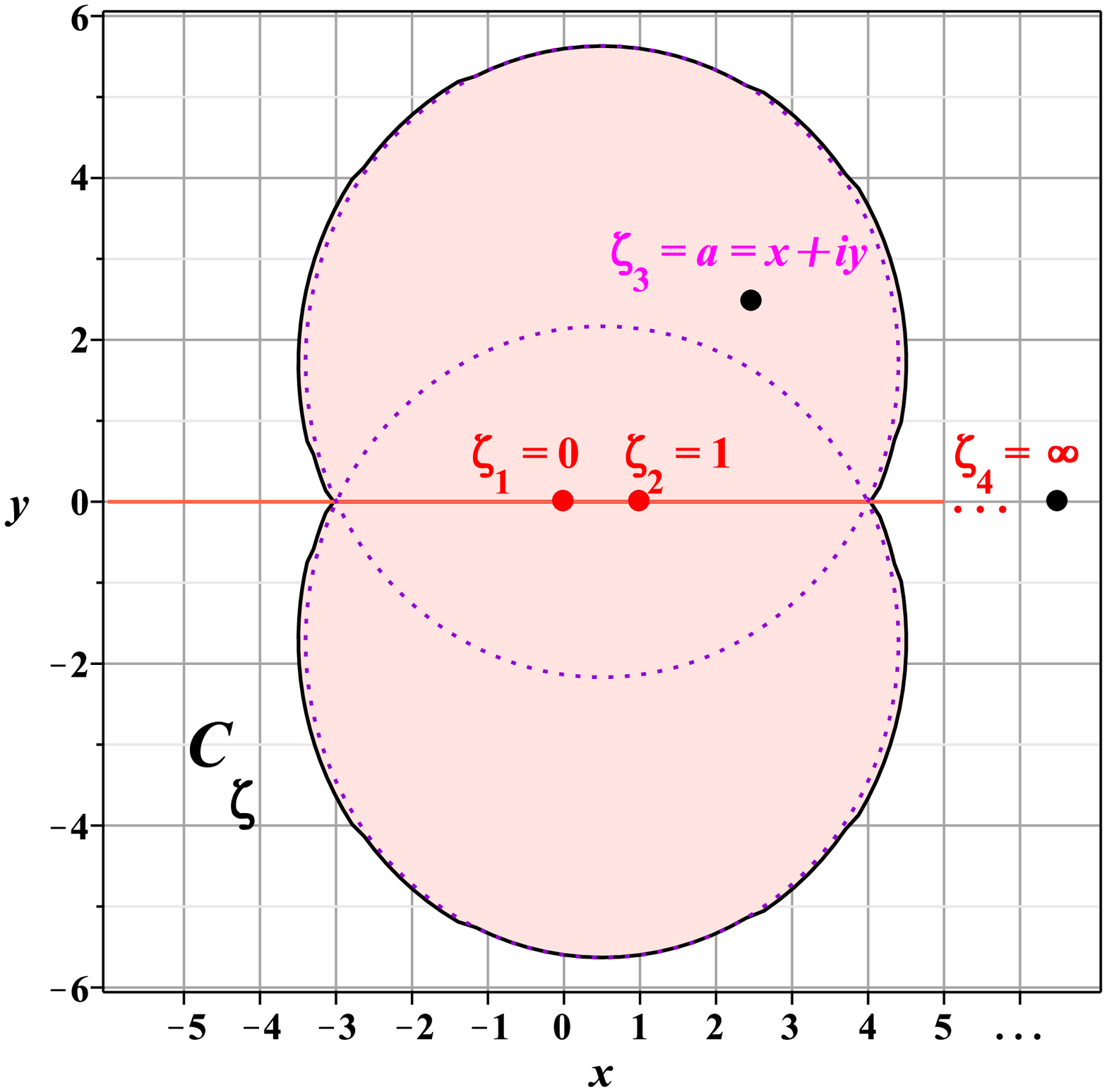}
\vskip .3truecm
\caption{The pink domain $\text{Dmn}(a)$ of the variable $a\in \mathbb{\tilde C}_\zeta$  with simultaneously valid \\Conditions A and B.}
\label{fig9}
\end{minipage}
\end{figure}

Since the Condition B also remains unchanged under previously used translations --
$\forall \phi$: $\Phi_1 \to \Phi_1+\phi, \Phi_2 \to \Phi_2+\phi, \Phi_3 \to \Phi_3+\phi$,
we can keep the fixing $\Phi_4=0$. Then we obtain on the torus $\mathbb{T}^{\,\,\,\,(2)}_{\Phi_1\Phi_2}$ different domains
of the validity of the Condition B, depending on the values of the parameter $a$. One example is shown in Fig.\ref{fig6}.
An example of a domain on the torus $\mathbb{T}^{\,\,\,\,(2)}_{\Phi_1\Phi_2}$ with the simultaneously valid Conditions A and B
is shown in Fig.\ref{fig7}. An example of a proper configuration in $\mathbb{\tilde C}_z$ is shown in Fig.\ref{fig7}. It turns out that such
configurations are possible only in a restricted domain $\text{Dmn}(a)\subset \mathbb{\tilde C}_\zeta$
of values of the parameter $a\in \mathbb{\tilde C}_\zeta$ - see Fig.\ref{fig9}.
The domain $\text{Dmn}(a)$ contains the union of two disks with the centers at points $\{1/2, \pm \sqrt{3}\}$ and radii $r\approx 3.9$,
but its exact form is more complicated and still unknown. Figure \ref{fig9} demonstrates the numerical result for the domain $\text{Dmn}(a)$.

\section{Some comments and concluding remarks}\la{Comments}

In the present paper, we represent a novel solution of the connection problem for local solutions to the general Heun's equation
based on the recently proposed symmetric form of this equation \cite{Fiziev14,Fiziev16}. Owing to the freedom to choose the positions
of the four regular singular points of the symmetric form of the general Heun's equation \eqref{dF} we are able to arrange those positions
in a way which ensures simultaneous well defined values of all local solutions at four special regular points $Z_{123},Z_{234},Z_{341},Z_{412}$
- the centers of the circles in the complex plane through each triple of singular points $z_{1,2,3},z_{2,3,4},z_{3,4,1},z_{4,1,2}$.
This permits us to find the coefficients of the connection matrices
in terms of local solutions at these regular points, see Eqs. \eqref{C:1,2,3,4}.
The special case when it turns possible to find the coefficients of the connection matrices
using only one regular point $Z_{klm}$ is also considered.

The final conclusion is that in contrast to the case of the hypergeometric function, in general, in the case of consideration
in the present paper it seems not possible to find
the coefficients of the connection matrices in terms of simpler functions than the general Heun's ones.
This problem needs more careful investigations since we still have no proof of nonexistence of some essential simplifications
of the coefficients \eqref{C:1,2,3,4} of the connection matrices due to the special character of the centers $Z_{klm}$
of the circles $C_{klm}$ where these coefficients are calculated.
(Note once more that the last regular points are completely determined by
Eq. \eqref{center} via the singular points $z_{1,2,3,4}$.)

\vskip .7truecm
\noindent{\bf \Large Acknowledgments}
\vskip .3truecm
The author is grateful to Professor Sergei Yu. Slavyanov for his helpful comments
and his kind encouragement,
to Professors Alexander Kazakov, Oleg Motygin and Artur Ishkhanyan, as well as
to other participants of the Section {\it Heun's equations and their applications}
of the Conference {\it Days on Diffraction 2016}, St. Petersburg, 27th of June- 01th of July, 2016
for their interest in the talk on the basic results of the present article
which were reported there for the first time.

Special thanks are also to the leader of the Maple-Soft-developers, Edgardo Cheb-Terrap,
for many useful discussions during the last years
on the properties of the Maple-Heun's functions and the computational problems with them.

The author is also thankful to the leadership of BLTP, JINR, Dubna for the support and good working conditions.
This article was also supported by the Sofia University Foundation {\it Theoretical and Computational Physics and Astrophysics}
and by the Bulgarian Agency for Nuclear Regulation, the 2014, 2015, 2016 grants.

%\setcounter{section}{1}

%\section*{References}

\end{document}